\documentclass[12pt,preprint]{aastex}
\newcommand{\etal}{{et al.}\@ }

\newcommand{\ie}{{i.e.,}\@ }

\newcommand{\Vmax}{$1/V_{\rm max}${ }}
\newcommand{\Teff}{$T_{\mathrm{eff}}$}
\newcommand{\logg}{{log $g$}}
\shorttitle{WDLF from SDSS Spectra}
\shortauthors{DeGennaro \etal}

\begin{document}

\title{White Dwarf Luminosity and Mass Functions from Sloan Digital Sky Survey Spectra}
\shorttitle{SDSS White Dwarf Luminosity Function}

\author{Steven DeGennaro\altaffilmark{1}, Ted von Hippel\altaffilmark{1}, D. E. Winget\altaffilmark{1},
S. O. Kepler\altaffilmark{2},  Atsuko Nitta\altaffilmark{3}, Detlev Koester\altaffilmark{4},
Leandro Althaus\altaffilmark{5,6}}
\altaffiltext{1}{The University of Texas at Austin, Department of Astronomy, 1 University Station C1400, Austin, TX 78712-0259}
\altaffiltext{2}{Instituto de F\'\i sica, Universidade Federal do Rio Grande do Sul, 91501-900 Porto-Alegre,RS,Brazil}
\altaffiltext{3}{Gemini Observatory, Hilo, HI 96720, USA}
\altaffiltext{4}{Institut f\"ur Theoretische Physik und Astrophysik, Universit\"at Kiel, 24098 Kiel, Germany}
\altaffiltext{5}{Facultad de Ciencias Astron\'omicas y Geof\'{\i}sicas, Universidad Nacional de La Plata, \
Paseo del Bosque S/N, 1900, La Plata, Argentina}
\altaffiltext{6}{Instituto de Astrof\'{\i}sica La Plata, IALP, CONICET}

\begin{abstract}
We present the first phase in our ongoing work to use Sloan Digital Sky
Survey (SDSS) data to create separate white dwarf (WD) luminosity functions 
for two or more different mass ranges.  In this paper, we determine 
the completeness of the SDSS spectroscopic white dwarf sample by 
comparing a proper-motion selected sample of WDs from SDSS imaging 
data with a large catalog of spectroscopically determined WDs.  We 
derive a selection probability as a function of a single color ($g-i$) 
and apparent magnitude ($g$) that covers the range $-1.0 < g-i < 0.2$ 
and $15 < g < 19.5$.  We address the observed upturn in log $g$ for 
white dwarfs with $T_{\mathrm{eff}} \lesssim$ 12,000K and offer 
arguments that the problem is limited to the line profiles and is 
not present in the continuum.  We offer an empirical method of 
removing the upturn, recovering a reasonable mass function for white 
dwarfs with $T_{\mathrm{eff}}<$ 12,000K.  Finally, we present a white 
dwarf luminosity function with nearly an order of magnitude (3,358) 
more spectroscopically confirmed white dwarfs than any previous work.
\end{abstract}


\section{Introduction}
Because white dwarfs cannot replenish the energy they radiate 
away---any residual nuclear burning is negligible and gravitational 
contraction is severely impeded by electron degeneracy---their 
luminosity decreases monotonically with time.  A thorough knowledge 
of the rate at which WDs cool can provide a valuable ``cosmic clock'' 
to determine the ages of many Galactic populations, including the 
disk \citep{Winget87,LDM,Leggett98,Knox99}, and open and globular 
clusters \citep{Claver95,vonHippel95,Richer98,Claver01,Hansen02,Hansen04,
vonHippel06,Hansen07,Jeffery07}.  With more accurate models of the cooling 
physics of white dwarfs, heavily constrained by empirical evidence, 
it may be possible to determine absolute ages with greater precision 
than using main-sequence evolution theory.  In addition to applications 
in astronomy, white dwarfs allow us to probe the physics of degenerate 
matter at temperatures and densities no terrestrial laboratory can 
duplicate.

Attempts at an empirical luminosity function (LF) for white dwarfs 
date as far back as \citet{Luyten58} and \citet{Weidemann67}.  The 
low-luminosity shortfall, discovered by \citet{Liebert79}, and 
attributed by \citet{Winget87} to the finite age of the Galactic 
disk, was confirmed and explored more fully when a greater volume 
of reliable data on low-luminosity WDs became available 
\citep{LDM, Wood92}.  More recently, Sloan Digital Sky 
Survey (SDSS) photometric data have been used to provide a much 
more detailed luminosity function with more than an order of 
magnitude more white dwarfs than previously attempted 
\citep{Harris06}, as well a new LF of a large sample of 
spectroscopically confirmed WDs \citep{Hu07}.  However, to 
date no one has published a well-populated luminosity function 
that does not include a wide range of masses and spectral types.  
Thus, much of the important physics of white dwarf cooling remains 
buried in the data.

Until recently, empirical WD luminosity functions, especially 
those derived from stars with spectra, have been hampered 
by a limited volume of reliable data.  This has forced a trade-off 
between the number of stars included in a sample, and their 
homogeneity; either a broad range of temperatures, masses, and 
spectral types must be used, or else the sample population of stars 
would be so small as to render reliable conclusions difficult.
Recently, the situation has changed dramatically.  Data from SDSS 
DR4 have yielded nearly 10,000 white dwarf spectra.  All of these 
spectra have been fitted with model atmospheres to determine their 
effective temperatures and surface gravities \citep{Kleinman04, 
Krzesinski04, Eisenstein06, Hugelmeyer06, Kepler07}.

In a companion paper to be published shortly, we intend to focus 
on how the WD cooling rate changes with WD mass.  
Theoretical work has been done in this area \citep{Wood92, 
Fontaine01}, but to date, attempts at creating an empirical LF to 
explore the effects of mass have relied on limited sample sizes 
\citep{Liebert05}.  In order to further isolate the effect of 
mass, we have chosen to study only the DA white dwarfs---white 
dwarfs which show only lines of hydrogen in their spectra---which 
comprise $\sim$86\% of all white dwarfs.

In addition to helping unlock the physics of white dwarfs, 
creating luminosity functions for several mass bins can also 
help to disentangle the effects of changes in cooling rates 
from changes in star formation rates.  A burst or dip in star 
formation at a given instant in Galactic history should be 
recorded in all of the luminosity functions, regardless of mass, 
and could be confirmed by its position across the various mass bins.  
For example, a short burst of increased star formation would be 
seen as a bump in each luminosity function, occurring at cooler 
temperatures in the higher mass LF (these stars, with shorter 
MS lifetimes, have had longer to cool).  On the other hand, 
features intrinsic to the cooling physics of the white 
dwarfs themselves should be seen in places that correspond 
with the underlying physics, which may be earlier, later, or
nearly concurrent across mass bins.  These effects include
neutrino cooling, crystallization, the onset of convective 
coupling \citep{Fontaine01}, and Debye cooling 
\citep{Althaus07}.

The current paper lays the groundwork for this analysis.
In Section \ref{data}, we introduce the data, examining the methods
used to classify spectra and derive quantities of interest (dominant
atmospheric element, \Teff, and log $g$).  We also address the observed
upturn in log $g$ for DAs below ${T}_{\mathrm{eff}} \sim$ 12,000K.  
We present several lines of reasoning that the upturn is an artifact
of the line fitting procedure, and propose an empirical method for
correcting the problem.  Section \ref{construction} outlines the
methods used to construct the luminosity and mass function and determine
error bars.

In Section \ref{completeness}, we present an analysis of the 
completeness of our data sample.  We use a well-defined sample 
of proper-motion selected, photometrically determined white 
dwarfs in SDSS \citep{Harris06} to determine our completeness
and derive a correction as a function of $g-i$ color and $g$ 
magnitude.  Finally, in Section \ref{discussion}, we present 
our best luminosity and mass functions for the entire DA 
spectroscopic sample and discuss the impact of both our 
empirical log $g$ correction and our completeness correction.

\section{The Data}
\label{data}
Our white dwarf data comes mainly from \citet{Eisenstein06}, a 
catalog of spectroscopically identified white dwarfs from the 
Fourth Data Release (DR4) of the Sloan Digital Sky Survey 
\citep{York00}.  The SDSS is a survey of $\sim$8,000 
square degrees of sky at high Galactic latitudes.  It is, first and 
foremost, a redshift survey of galaxies and quasars.  Large ``stripes'' 
of sky are imaged in 5 bands (u,g,r,i,z) and objects are selected, 
on the basis of color and morphology, to be followed up with 
spectroscopy, accomplished by means of twin fiber-fed spectrographs, 
each with separate red and blue channels with a combined wavelength 
coverage of about 3800 to 9200\AA{ }and a resolution of 1800.  Objects 
are assigned fibers based on their priority in accomplishing SDSS science 
objectives, with high redshift galaxies, ``bright red galaxies'' and 
quasars receiving the highest priority.  Stars are assigned fibers for 
spectrophotometric calibration, and other classes of objects are only 
assigned fibers that are left over on each plate.  More detailed 
descriptions of the target selection and tiling algorithms can be found 
in \citet{Stoughton02} and \citet{Blanton03}.

Though white dwarfs are given their own (low priority) category in the 
spectroscopic selection algorithms, very few white dwarfs are targeted 
in this way.  Rather, most of the white dwarfs in SDSS obtain spectra 
only through the ``back door,'' most often when the imaging pipeline mistakes
them for quasars.  \citet{Kleinman04} list the various 
algorithms that target objects ultimately determined to be white 
dwarfs in DR1 (their Table 1).  White dwarfs are most commonly targeted 
by the QSO and SERENDIPITY\_BLUE algorithms, with significant contributions 
also from  HOT\_STANDARD (standard stars targeted for spectrophotometric 
calibration) and SERENDIPITY\_DISTANT.  Of the significant contributors, 
the STAR\_WHITE\_DWARF category contributes the least to the 
population of WD spectra.

The SDSS Data Release 4 contains nearly 850,000 spectra.  
Several groups have already attempted to sort through them to find 
white dwarfs: \citet{Harris03} for the Early Data Release, 
\citet{Kleinman04} for Data Release 1 (DR1), and most recently, 
\citet{Eisenstein06} for the DR4, from which the majority of our 
data sample derives, though a handful of stars from DR1 omitted by 
Eisenstein have been re-included from \citet{Kleinman04}.  Most 
recently, \citet{Kepler07} have refit the DA and DB stars from 
\citet{Eisenstein06} with an expanded grid of models.  A complete 
analysis of the methods by which candidate objects are chosen, spectra 
fitted, and quantities of interest are calculated can be found in 
\citet{Kleinman04}, \citet{Eisenstein06}, and \citet{Kepler07}.  
We put forth a brief outline here, with special attention paid to those 
aspects important to our own analysis.

Objects in the SDSS spectroscopic database are put through several 
cuts in color designed to separate the WDs from the main stellar locus.  
Figure 1 in \citet{Eisenstein06} shows the location of these cuts.  
The chief failing of their particular choices of cuts, as noted by the 
authors, is that WDs with temperatures below $\sim$8,000K begin to 
overlap in color-color space with the far more numerous A and F stars, 
and they have not attempted to dig these stars out.  The SDSS 
spectroscopic pipeline calculates a redshift for each object by 
looking for prominent lines in the spectrum.  Objects with redshifts 
higher than z=0.003 are eliminated, unless the object has a proper 
motion from USNO-A greater than 0.3'' per year.  Because the 
spectroscopic pipeline is fully automated, occasionally DC white 
dwarfs show weak noise features that can be misinterpreted as 
low-confidence redshifts.  Other types of WD, particularly magnetic 
WDs, can fool the pipeline as well.  In the present paper we are 
concerned chiefly with DA white dwarfs, so this incompleteness is 
of importance only insofar as we use the entire set of white dwarf 
spectral types to derive our completeness correction, as outlined 
in Section \ref{completeness}.  We explore the implications of this 
more fully in that section.

\citet{Eisenstein06} then use a $\chi^{2}$ minimization technique 
to fit the spectra and photometry of the candidate objects with 
separate model atmospheres of pure hydrogen and pure helium 
\citep{Finley97, Koester01} to determine the dominant element, 
effective temperature, surface gravity, and associated errors.  
As their Figure 2 demonstrates, they recover a remarkably complete 
and uncontaminated sample of the candidate stars.  They 
believe that they have recovered nearly all of the DA white dwarfs hotter 
than 10,000K with SDSS spectra.

These stars form the core of our data sample.  Their final table lists 
data on 10,088 white dwarfs.  Of these, 7,755 are classified as single, 
non-magnetic DAs.  \citet{Kepler07} re-fit the spectra for these 
stars using the same autofit method and Koester model atmospheres, 
but with a denser grid which also included models up 
to log $g$ of 10.0.  Where they differ from Eisenstein's, 
we use these newer fits in our analysis.  Of these 7,755 entries, 
$\sim$600 are actually duplicate spectra of the same star.  For our 
analysis we take an average of the values derived from each 
individual spectra weighted by the quoted errors.  Our final sample 
contains 7,128 single, non-magnetic DA white dwarfs.

\begin{figure}[!tp]
\includegraphics[angle=270,width=\textwidth]{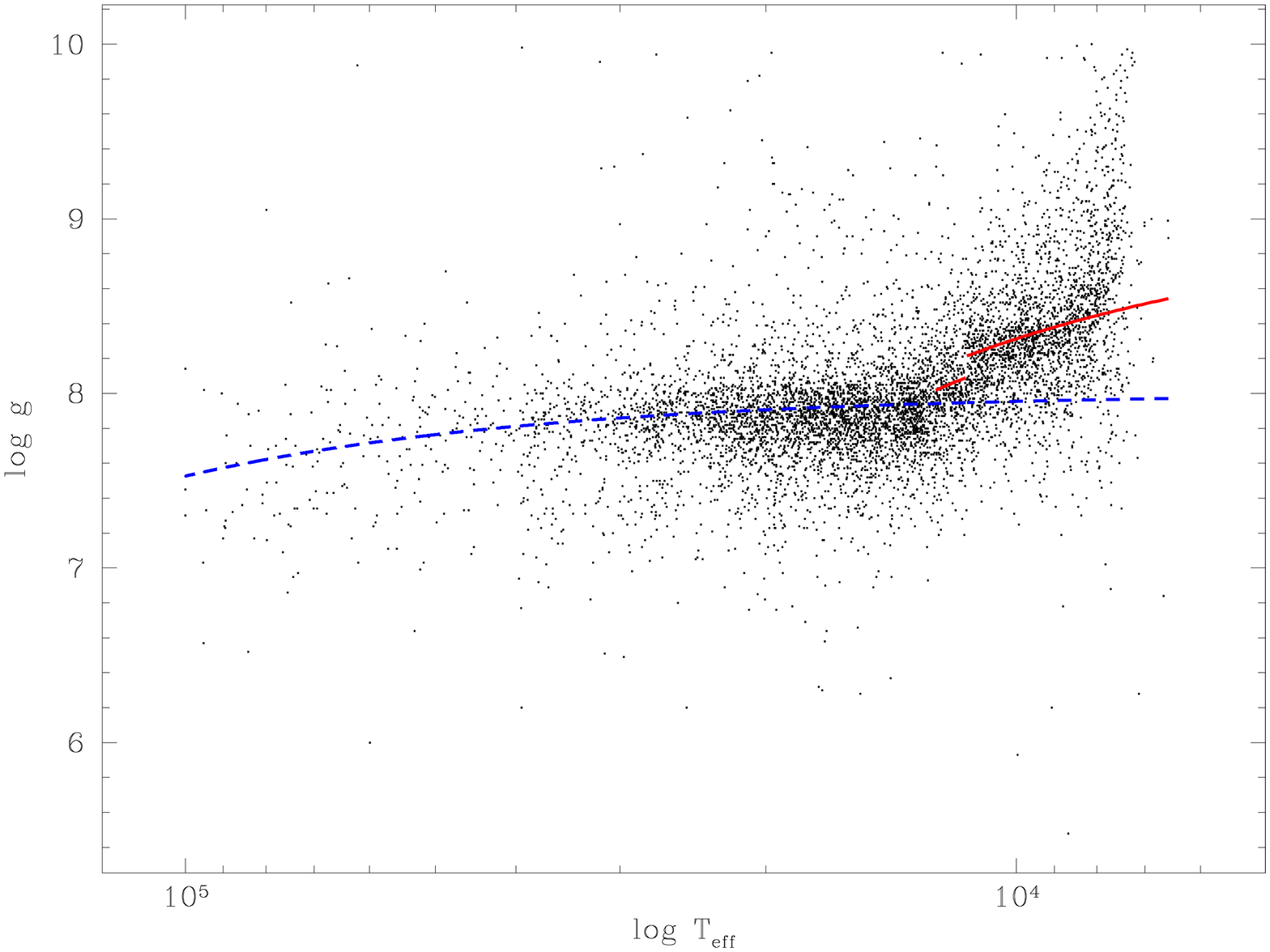}
\caption{log $g$ v. log ${T}_\mathrm{{eff}}$ for the white 
dwarfs in our sample.  At temperatures below $\sim$12,500K, the log $g$ 
values begin to rise to an extent unexplained by current theory.  The 
solid line is a function empirically fit to the real data.  The dashed 
line is the modest rise predicted by theory.  The excess at a given \Teff{ }is 
subtracted from the measured log $g$ value for some of our luminosity 
functions.}
\label{upturn}
\end{figure}

As noted by \citet{Kleinman04} and others, the surface gravities 
determined from Sloan spectra show a suspicious upturn below 
temperatures of about 12,000K which increases at cooler temperatures, 
as shown in our Figure \ref{upturn}. 

A number of separate pieces of evidence argue that this upturn 
in log $g$---and thus mass---is an artifact of the models and not
a real effect.  Not least among these is that no one has yet 
provided any satisfactory mechanism by which WDs could gain 
enough mass or shrink enough in radius as they cool to 
account for the magnitude of the effect.  We do expect a slight
increase in mass at cooler temperatures because in a galaxy of
finite age, the cooler white dwarfs must come from higher
mass progenitors.  This is the reason for the upward slope of the 
blue dashed line in Figure \ref{upturn}.  However, this effect 
is clearly small compared to the upturn observed in the actual data.

Furthermore, \citet{Engelbrecht06}, and \citet{Kepler07} 
demonstrated that the masses derived solely from the colors do not 
show an increase in mass for cooler stars, which indicates that 
the problem is not physical, but a result of either the line 
fitting procedure or the line profiles themselves.

\begin{figure}[!tp]
\includegraphics[angle=270,width=\textwidth]{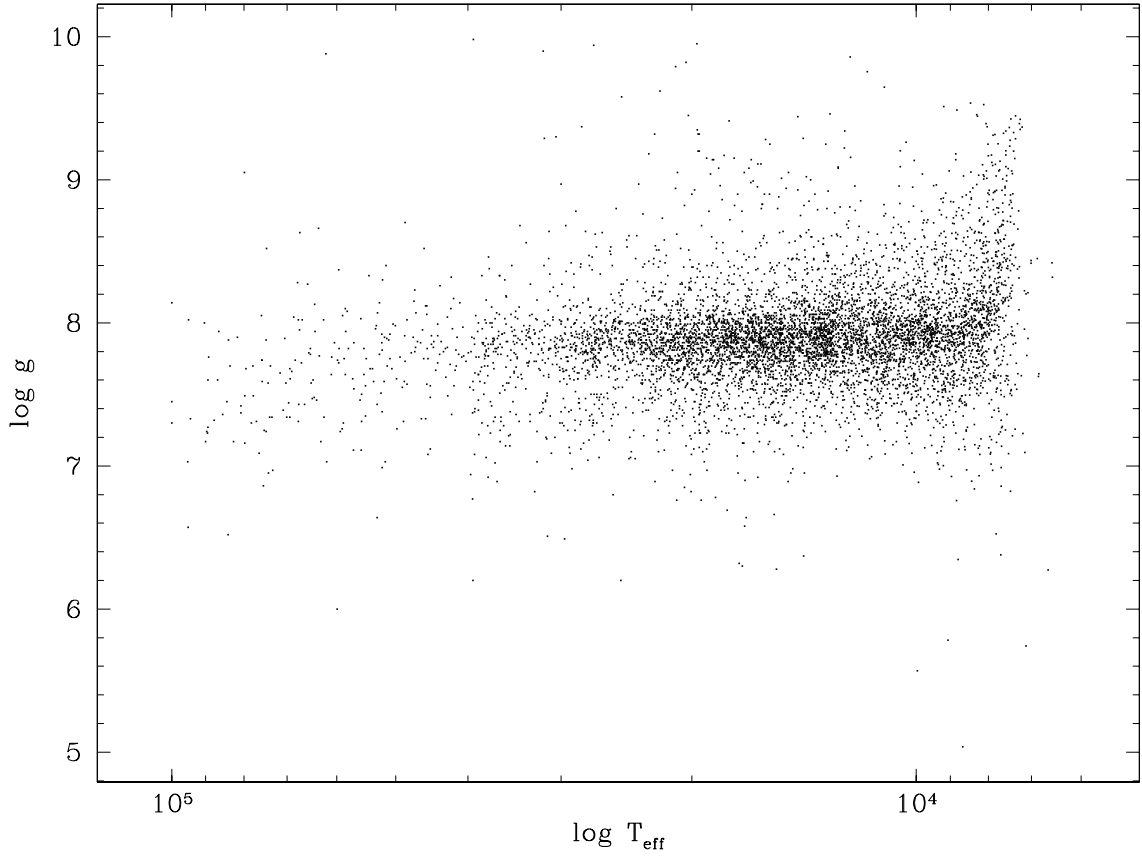}
\caption{log $g$ v. log ${T}_\mathrm{{eff}}$ with the upturn 
in log $g$ removed.}
\label{noupturn}
\end{figure}

\begin{figure}
\includegraphics[width=6.5in]{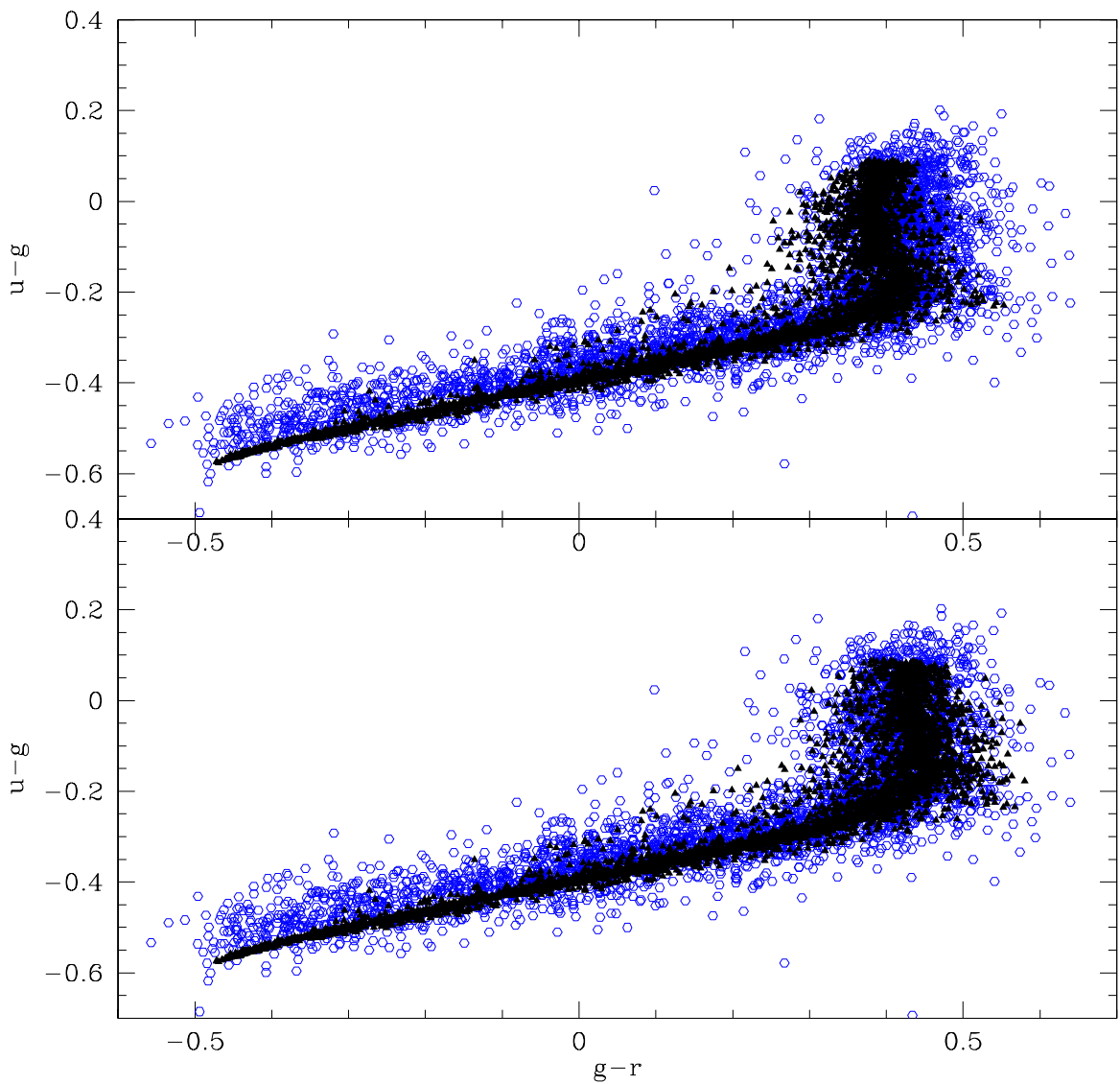}
\caption{A comparison of the theoretical colors of the SDSS 
WDs, derived from the atmospheric fits (black triangles), with the 
observed colors, as measured by the SDSS photometry (open blue circles).  
In the upper panel, the colors of the model atmospheres do not agree 
with the observed colors at low temperatures, indicating a problem with 
the line fitting for stars cooler than $\sim$ 12,500K.  In the lower panel, 
where the excess log $g$ has been removed, the colors agree much better.}
\label{colors}
\end{figure}

Figure \ref{colors} further illustrates the above point.  The upper
panel shows the colors derived from the synthetic spectra at the 
values of \Teff{ }and log $g$ quoted by \citet{Kepler07} (\ie the 
values in Figure \ref{upturn}), overlaid on the actual SDSS 
photometry for the same objects.  Contrast this with the lower panel, 
which instead shows the colors derived from the synthetic spectra 
when the excess log $g$ has been removed (in a manner described below; 
the resulting values are shown in Figure \ref{noupturn}).  The colors 
in the latter figure agree much better with the measured color of 
the object.

Furthermore, \citet{Kepler07} found a similar increase 
in mean mass for the SDSS DB white dwarfs below 
{${T}_{\mathrm{eff}} \sim$ 16,000K}.  They conclude that since 
a) the problem only shows up in the line profiles and not 
the continuum, and b) the onset of the effect in both hydrogen 
(DA) and helium (DB) atmosphere WDs occurs at just the effective 
temperature where the neutral species of the atmospheric element 
begins to dominate, then the problem lies in the treatment of 
line broadening by neutral particles.  This is supported further 
by the fact that as the species continues to become more 
neutral (\ie as the temperature drops), the problem grows worse.

However, more recent model calculations indicate that neutral 
broadening is not important in the DA white dwarfs at temperatures 
down to at least than 8,500K.  Other possible mechanisms to explain 
the observed upturn in log $g$ include a flawed or incomplete
treatment of convection, leading to errors in the temperature 
structure of the outer layers of the WD models, or the
convective mixing of helium from a lower layer in the atmosphere 
\citep{Bergeron90, Bergeron95b}.  The latter would require a hydrogen 
layer much thinner than any seismologically determined in a DA
so far \citep{Bradley98, Bradley01, Bradley06}.

Until the problem with the model atmospheres is resolved, the best
we can do is to empirically remove the log $g$ upturn.  For a given 
\Teff, we subtract the excess in the measured mean value (as fit by 
the red solid lines in Figure \ref{upturn}) over the theoretically 
expected mean (blue dashed line).  Figure \ref{noupturn} shows the 
resulting values used.  In fitting out the upturn this way,
we make two implicit assumptions.  First, that the excess log $g$ is
a function only of \Teff;  if the problem is indeed due to the 
treatment of neutral particles, we would expect only a small 
dependence on log $g$.  Second, we assume that the problem affects 
only the log $g$ determination and not \Teff.  This latter assumption 
is unlikely to be true, as the two parameters are correlated.  
In Section \ref{discussion} we explore more fully the impact of 
this fitting procedure on the luminosity and mass functions.

\section{Constructing The Luminosity And Mass Functions}
\label{construction}

Since we are dealing with a magnitude-limited sample, the most 
luminous stars in our sample can be seen to much further distances 
than the intrinsically fainter stars.  We thus expect more of them,
proportionally, than we would in a purely volume-limited sample, and
must make a correction for the different observing volumes.  As
shown by \citet{Wood98} and \citet{Geijo06}, the \Vmax 
method of \citet{Schmidt68} \citep[described more fully in, e.g.][]{Green80,FLG} 
provides an unbiased and reliable characterization of the WDLF.

In the \Vmax method, each star's contribution to the total space density is 
weighted in inverse proportion to the total volume over which it 
would still be included in the magnitude limited sample.  Since the 
stars are not spherically distributed, but lie preferentially in the 
plane of the Galaxy, an additional correction for the scale height of 
the Galactic disk must be included.  For the purposes of comparison 
with previous work, we adopt a scale height of 250pc.

To determine the absolute magnitude of each WD, we use the effective 
temperatures and log $g$ values provided by \citet{Kepler07}---as 
corrected in Section \ref{data}---and fit each WD with an
evolutionary model to determine the mass and radius.  For 
{$7.0 <$ log $g < 9.0$}, we use the mixed C/O models of \citet{Wood95} 
and \citet{Fontaine01}, as calculated by \citet{Bergeron95a}.  For 
{$9.0 <$ log $g < 10.0$}, we use the models of \citet{Althaus05}
with O/Ne cores, including additional sequences for masses larger than 
1.3 ${M}_{\odot}$ calculated specifically for \citet{Kepler07}.  Once we 
know the radius, we can calculate the absolute magnitude in each Sloan 
band by convolving the synthetic WD atmospheres of Koester \citep{Finley97, 
Koester01} with the Sloan filter curves.  We apply bolometric 
corrections from \citet{Bergeron95a} to determine the bolometric 
magnitude.  For the handful of stars ($\sim$80-100) with log $g$ values 
outside the range covered by Bergeron's tables, we use a simple 
linear extrapolation.

We then determine photometric distances to each star from the 
observed SDSS $g$ magnitude.  SDSS, being concerned mostly with 
extragalactic objects, reports the total interstellar absorption 
along each line of sight from the reddening maps of \citet{Schlegel98}.  
Since the objects in our sample lie within the Galaxy, and most of 
them within a few hundred parsecs, they are affected by only a 
fraction of this reddening.  Following \citet{Harris06}, we 
therefore assume: 1) that objects within 100pc are not affected 
by reddening, 2) objects with Galactic height $\left|{z}\right|$ $>$ 
250pc are reddened by the full amount, and 3) that the reddening 
varies linearly between these two values.  The distances and 
reddening are then fit iteratively from the observed and calculated 
absolute $g$ magnitude.  In practice, the reddening correction 
makes very little difference to the final LF (typical ${A}_{g}$ values
range from 0.01 to 0.05).

We calculate error bars on the luminosity function using a Monte-Carlo
simulation, drawing random deviates in \Teff, \logg, and each band of
photometry from gaussian distributions centered around the measured 
value.  The standard deviations in \Teff{ }and log $g$ we use for 
this scattering are 1.2 times the formal errors quoted in 
\citet{Eisenstein06} (their own analysis, based on repeated autofit 
measurements on duplicate spectra of the same stars, suggests that 
the formal errors derived by their method are $\sim$20\% too small).  
The photometry errors come directly from the SDSS database.  After 
scattering the parameters in this way, we recalculate the LF.  We then 
add in quadrature the standard deviation of each LF bin after 200 
iterations and the counting error for each bin (the errors for each 
individual star---taken to be of the order of the star's \Vmax 
statistical weight---summed in quadrature).
  
At a S/N of 16---the mean for the stars in our sample brighter than 
$g = 19.5$---formal errors in \Teff{ }and log $g$ are of order 1.5\%.  
When propagated through our code, the mean errors in ${M}_{bol}$ and 
mass are 0.35 dex and 9\% (0.05 ${M}_{\odot}$) respectively.  For the 
stars brighter than $g = 19.0$ used to compile our mass functions 
the average S/N is 19.5, leading to errors in ${M}_{bol}$ and mass 
of 0.35 dex and 7\% (0.04 ${M}_{\odot}$).

\section{Completeness Corrections}
\label{completeness}

The chief difficulty we have encountered in deriving our luminosity 
functions is unraveling the complicated way in which SDSS objects 
are assigned spectral fibers.  SDSS is foremost a survey of 
extragalactic objects and rarely targets white dwarfs for 
follow up spectroscopy explicitly.  Most of the objects in our sample 
are targeted by some other algorithm.  In particular, there is 
considerable overlap in color between white dwarfs and many QSOs.

A completeness correction could, in theory, be built from ``first principles.''  
We know, for each object in the SDSS spectroscopic database, by which 
algorithm(s) it was targeted (or rejected) for spectroscopy, and by which 
algorithm it was ultimately assigned a fiber.  And for each algorithm, 
we know which objects were targeted, which were ultimately assigned a 
fiber, and which, of the targeted objects, turned out to be WDs.  However, 
the selection process is a multi-variate function of 5 apparent magnitudes, 
and colors in spaces of as many as 4 dimensions (which vary based on the 
algorithm), as well as the complex tiling algorithm.  We believe such an
undertaking to be unnecessary.  Instead we have chosen to compare our 
sample with the stars used to derive the WDLF of \citet{Harris06}.  
Given certain assumptions about completeness and contamination in both 
data sets, we derive a completeness correction as a function of a single 
color index {($g - i$)} and $g$ magnitude.

The \citet{Harris06} sample comes from photometric data in the 
SDSS Data Release 3.  They selected objects by using the reduced proper 
motion diagram to separate WDs from more luminous subdwarfs of the 
same color.  Briefly, they used color and proper motion 
\citep[from USNO-B][]{Munn04} to determine WD candidates from SDSS 
imaging data.  They then fit candidates with WD model atmosphere colors 
to determine temperatures and absolute magnitudes, from which they derived 
photometric distances and---together with proper motion---tangential 
velocities.  In order to minimize contamination, they adopted a tangential 
velocity cutoff of 30 km/s and rejected all stars below this limit.  The 
remaining 6,000 objects are, with a high and well-defined degree of 
certainty ($\sim98 - 99\%$), likely to be white dwarfs.

If the database of SDSS spectra were complete, all of these objects would 
(eventually) have spectra, and all but the 1-2\% of contaminating objects 
would be confirmed to be WDs.  Furthermore, all of the WDs that did 
\emph{not} make it into the Harris \etal sample---because they were either 
missing from the \citet{Munn04} proper motion catalog, or had a tangential 
velocity below 30 km/s---would also all have spectra.  In such a perfect 
world, of course, no completeness correction would be necessary.  However, 
since SDSS does not obtain a spectrum of every object in its photometric 
database, a significant percentage of the objects in Harris \etal will not 
have spectra, or else will be dropped at some later point by Eisenstein \etal 
and thus not make it into our spectroscopic sample.  Our goal, then, is to 
look at all of the WDs in the Harris \etal sample that potentially \emph{could} 
have made it into our sample, and determine which ones in fact did.  If we 
assume that the WDs \emph{not} in Harris \etal follow the same distribution 
(an assumption we discuss more fully below), then we can take this as a 
measure of the overall detection probability and invert it to get a 
completeness correction.

The imaging area of the DR3, from which Harris \etal derive their sample, 
is not the same as the spectroscopic area in the DR4.  Therefore, for the 
purposes of this comparison, we removed all stars not found in the area 
of sky common to the two data sets from their respective samples.  This 
left 5,340 objects classified as white dwarfs by Harris \etal that could 
potentially have been recovered by Eisenstein et al.  Of these, 2,572 were 
assigned spectral fibers in DR4, and 2,346 were ultimately confirmed by 
Eisenstein \etal to be white dwarfs.

Since we wish to restrict our analysis to single (\ie non-binary) DA white 
dwarfs, we removed all stars classified as DA+M stars in either catalog.  
Unfortunately, given that the Harris catalog contains no further 
information as to the type of WD, we were unable to remove the non 
DA stars and simply compare what remains with the Eisenstein sample.  
Instead, we compute the completeness for all of the WDs, under the 
assumption---explored more fully below---that DAs, as the largest 
component of the WD population, dominate the selection function.

\begin{figure}[!tp]
\includegraphics[angle=270,width=\textwidth]{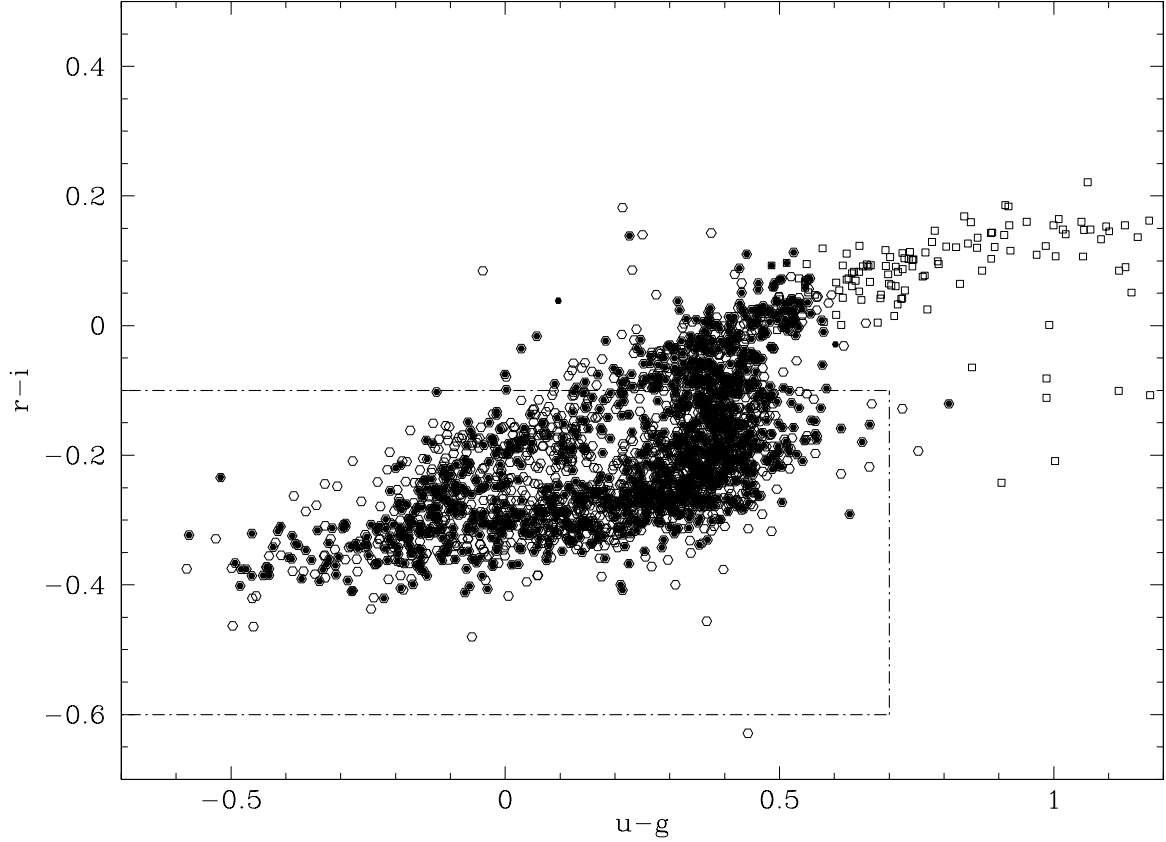}
\caption{Color-color plot of the white dwarfs in the two samples 
used to derive our completeness correction. Open symbols are WDs from the Harris 
\etal (2006) sample that a) were in the area of sky covered by spectroscopy in 
DR4, b) had ${V}_{tan} \ge 30$km/s, and c) were not determined by $i$- and 
$z$-band excess to be WD + main-sequence binaries.  The filled circles 
are the stars for which SDSS obtained spectra and \citet{Eisenstein06} confirmed 
to be WDs.  The dashed box shows a two-dimensional projection of the QSO 
targeting algorithm's exclusion region.  The open gray squares are the WDs from 
Harris \etal that lie outside Eisenstein et al.'s color-color cuts.  For clarity,
only half of the points have been plotted.}
\label{QSOfig}
\end{figure}

Figure \ref{QSOfig} shows a comparison of the two samples.  The open 
symbols are the complete Harris \etal sample (excluding those, as 
mentioned above, with {${V}_{tan} < 30$ km/s}, those not in the region of 
sky covered by spectroscopy, and the DA+M stars).  The gray squares lie 
outside the cuts in color-color space imposed by Eisenstein {et al.}  
They may have spectra in SDSS, but they were not fit by Eisenstein {et al.}, 
and therefore will not have made it into our sample.  The filled green circles 
are the stars that \emph{are} in Eisenstein \etal{ }In other words, if the
SDSS spectral coverage of WDs were complete, and Eisenstein \etal recovered
every WD spectra in SDSS, then all of the open circles would be filled.  The 
inside of the blue box is the exclusion region for SDSS's QSO targeting 
algorithm \citep{Richards02}, specifically implemented to eliminate WDs 
from their sample.  Note that our sample is more complete for the stars outside 
this region.

\begin{figure}[!tp]
\includegraphics[angle=270,width=\textwidth]{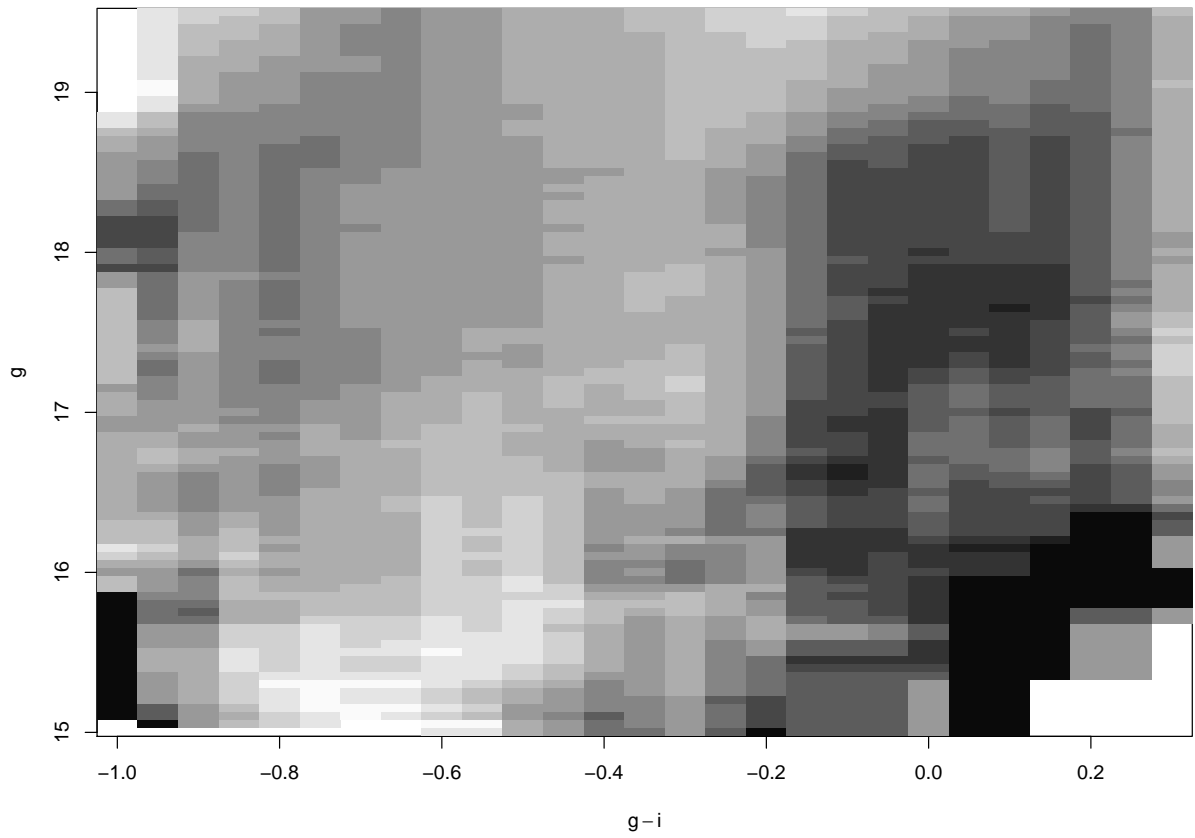}
\caption{A map of our completeness correction.  Darker 
areas indicate more complete regions of the figure, with black being 100\% 
complete.  The overall completeness is of order $\sim$50\%.}
\label{completenessfig}
\end{figure}

Figure \ref{completenessfig} shows the discovery probability as a function of 
$g-i$ color and $g$ magnitude.  Darker areas mean a higher probability of 
discovery, with black indicating that all the WDs in the Harris \etal 
sample in that area of color-magnitude space made it into our sample.  
We have performed a box smoothing to eliminate small scale fluctuations.

There is a drop off in discovery probability for stars bluer than 
${g-i\sim -0.2}$ at all apparent magnitudes.  This corresponds to the 
red edge of the exclusion region of the QSO targeting algorithm, as noted above.
The QSO algorithm is also itself a function of apparent magnitude, which 
accounts for the general decrease at fainter magnitudes in the 
red half of the diagram, and the much steeper drop off between ${g\simeq19}$
and ${g\simeq19.5}$.  The bluer stars (${g-i\lesssim -0.2}$), most of which 
are targeted by the HOT\_STANDARD or SERENDIPITY\_BLUE algorithms, 
show the opposite: a slight increase at fainter magnitudes.

\begin{figure}[!tp]
\includegraphics[angle=270,width=\textwidth]{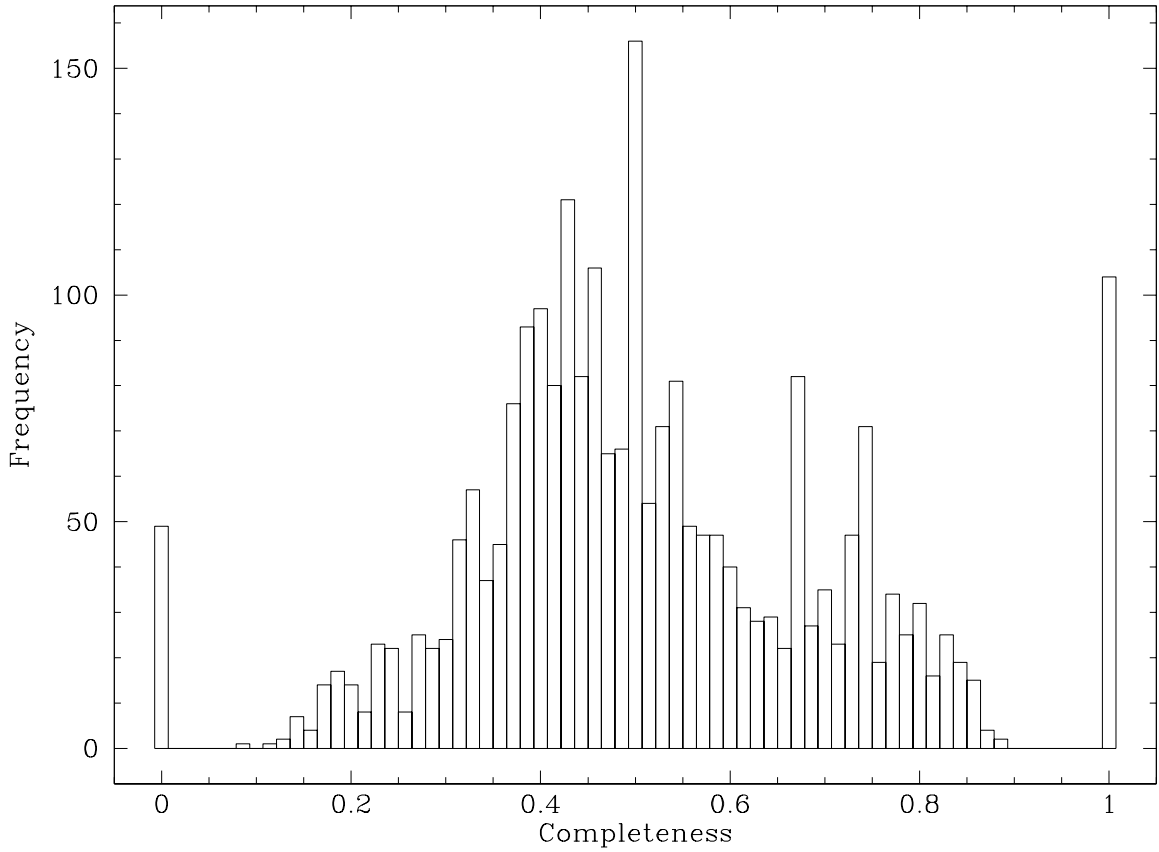}
\caption{A histogram of the completeness values in Figure 
\ref{completenessfig}.  Most of the 0, 1, and 0.5 values come from color-magnitude 
regions in which there are only one or two stars in the Harris \etal sample available 
for comparison.}
\label{completenesshist}
\end{figure}

To give a better sense of the order of magnitude of our completeness, 
Figure \ref{completenesshist} shows a histogram of the values in 
Figure \ref{completenessfig}.  For most of the cells that end up in 
the bins for 0, 1, and 0.5, the Harris \etal sample contains only one 
or two stars.  The mean completeness for the whole sample is $\sim 51\%$.

To derive our final completeness correction, we must further consider the 
incompleteness and contamination in the Harris \etal sample itself.  
Assuming that the SDSS photometric database is essentially complete 
down to $g=19.5$, then the incompleteness in Harris \etal comes mainly from 
two sources: 1) the incompleteness in the \citet{Munn04} proper motion 
catalog, and 2) the tangential velocity limit of 30 km/s imposed, which 
results in some low tangential velocity WDs being dropped from the sample. 
However, with one negligible exception, none of the criteria used to target 
objects for spectroscopy in SDSS, nor those used by Eisenstein \etal to 
select white dwarf candidates, depends explicitly on proper motion or 
tangential velocity.  Thus we assume that the low-velocity stars---dropped 
from the Harris \etal sample---will be recovered by Eisenstein with the 
same probability as the high-velocity stars---\ie the stars in Figure 
\ref{QSOfig}.

Contamination poses a bit more challenging problem.  At first glance, 
it would seem that the reverse of the above process could be applied, 
whereby those objects in Harris \etal which did get spectral 
fibers---but were ultimately rejected as WDs by Eisenstein {et al.}---could 
be removed from the sample, and those that did \emph{not} 
get spectra could be assumed to follow the same distribution.  This latter 
assumption, however, is unlikely to be true.  SDSS gives very low priority 
to targeting white dwarfs specifically, and we would thus expect a larger 
fraction of the objects that get spectral fibers to turn out to be 
contaminating objects (in particular QSOs, of which we found 13 in the 
Harris \etal sample) than if the fibers were assigned purely randomly.
Furthermore, many of the 225 objects which have spectra in DR4 but are 
not included in the Eisenstein catalog may actually be white dwarfs 
which Eisenstein's algorithms dropped for some other reason, e.g. they 
lie outside the color and magnitude ranges used for initial candidate 
selection, or there is a problem (low S/N, bad pixels) with the spectrum.  
Approximately 100 appear to be DC white dwarfs to which the SDSS 
spectroscopic pipeline assigned erroneous redhifts on the basis of weak 
noise features.  Ultimately, we have chosen to adopt the contamination 
fraction of Harris \etal (2\%) for the whole sample, and have reduced 
our final completeness correction accordingly.  This choice has a 
negligible effect on the small scale structure of the WDLF in which 
we are interested.

Finally, we note that the Harris \etal sample has an apparent magnitude 
limit of ${g=19.5}$, whereas the spectroscopic sample contains objects 
down to ${g\simeq20.5}$.  Given that the SDSS targeting algorithms are 
themselves functions of apparent magnitude, our completeness correction 
is as well.  An extrapolation of our discovery probability is problematic 
in this area, though, because this is just the apparent magnitude where 
the QSO targeting algorithm drops off rapidly.  We have decided to impose 
a magnitude cutoff of {$g=19.5$} in our sample.  This reduces our 
sample by nearly a half, with a corresponding increase in counting error.  
However, because SDSS spectra have a small range of exposure times 
(45-60min), fainter apparent magnitude usually translates directly 
into lower S/N and larger errors in derived parameters.

\begin{figure}[!tp]
\includegraphics[angle=270,width=\textwidth]{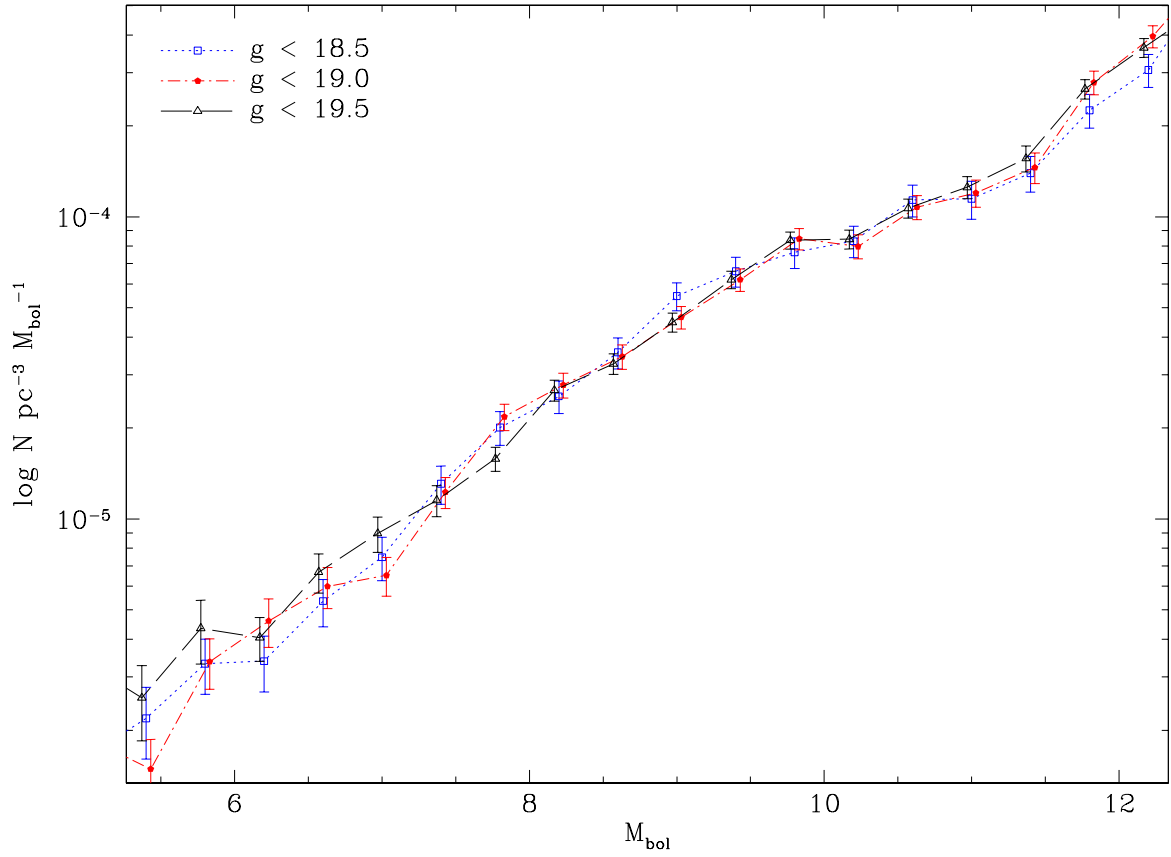}
\caption{Luminosity functions for three different limiting
magnitudes.  We take the good agreement between the curves to indicate that 
our completeness correction (and the \Vmax correction) are working properly.}
\label{mlimcomp}
\end{figure}

Figure \ref{mlimcomp} shows the luminosity functions we derive for 
different choices of limiting magnitude.  We take the generally good
agreement between the curves to indicate that our completeness correction
is doing its job correctly in the $g$ magnitude direction.

\begin{figure}[!tp]
\includegraphics[angle=270,width=\textwidth]{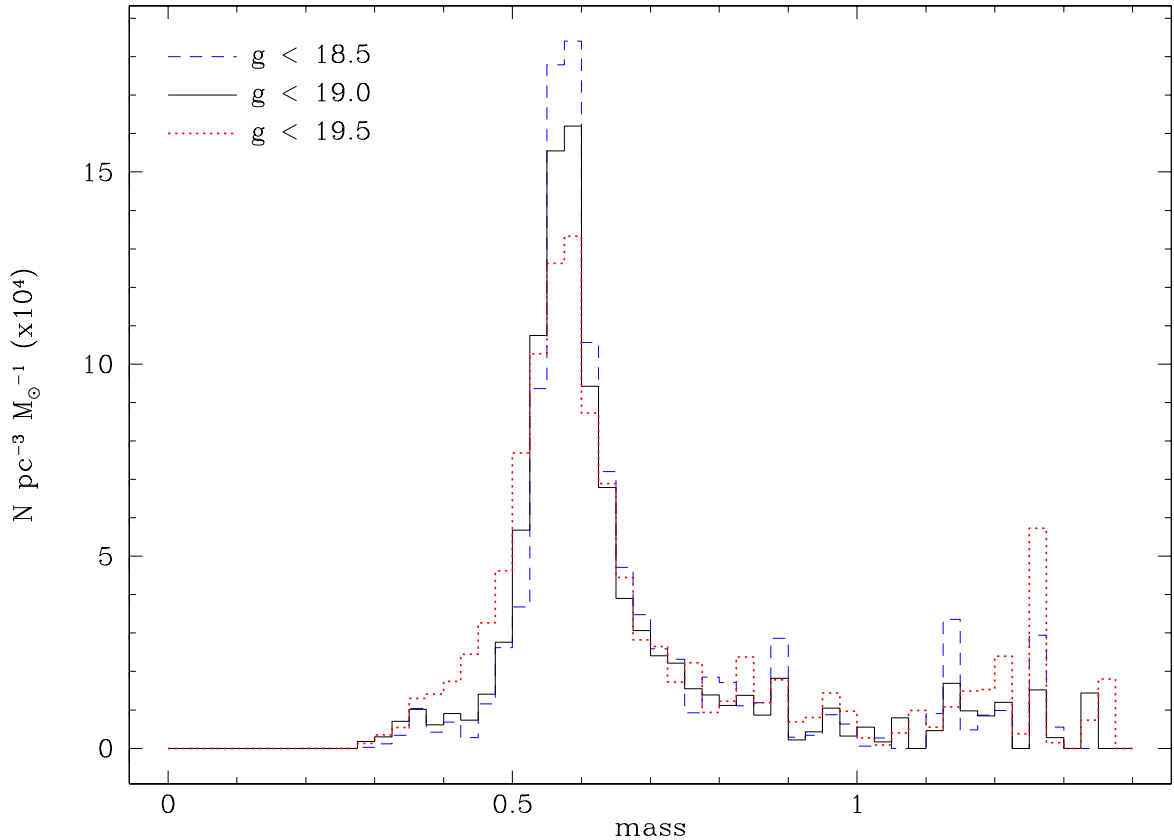}
\caption{Mass functions for three different limiting
magnitudes.  Because of the essentially fixed integration time for SDSS spectra, 
objects with fainter apparent magnitudes generally have lower signal to noise, 
which translates directly into larger uncertainties in the derived parameters 
(\Teff, \logg, and mass).  Hence, as we include stars with fainter apparent 
magnitudes, more stars scatter out of the peak, broadening the mass function.}
\label{mlimmasscomp}
\end{figure}

Figure \ref{mlimmasscomp} similarly shows the mass functions we derive for 
different choices of limiting magnitude.  In the case of the mass function, 
the S/N of the spectra becomes a much bigger factor.  As a consequence of the
essentially constant exposure times of SDSS spectra, the parameters 
(\Teff{ }and \logg) determined from the spectra of fainter objects 
have larger errors, which causes a larger error in mass.  Thus, the MF is 
broadened when stars with ${g > 19.0}$ are included.  For this reason, 
\citet{Kepler07} limited their mass functions to stars with ${g \leq 19.0}$, 
and we follow their lead for the remaining MFs in the current paper.

\section{Luminosity Functions And Discussion}
\label{discussion}

\begin{figure}[!tp]
\includegraphics[angle=270,width=\textwidth]{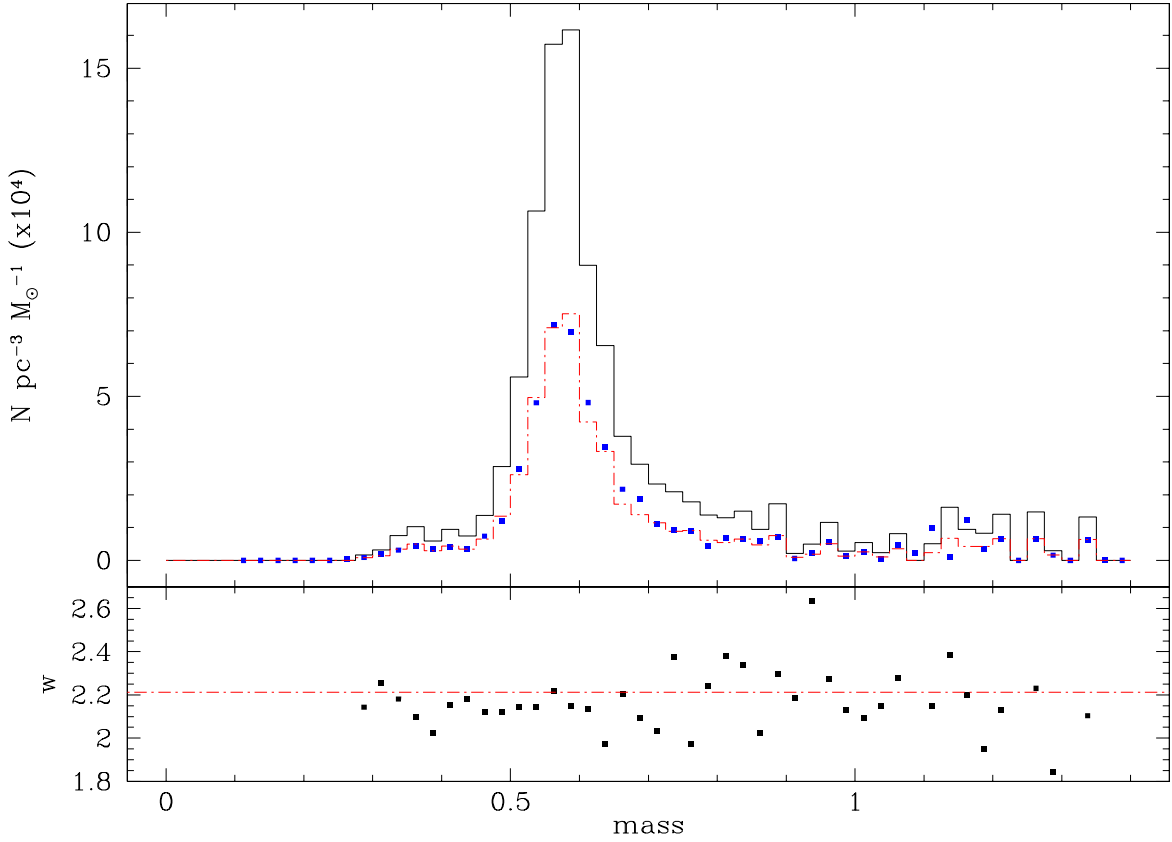}
\caption{The white dwarf mass function for all WDs with 
${T}_{\mathrm{eff}} >$ 12,000K and $g < 19.0$. The dashed line in the upper 
panel is the MF corrected only for {$1/V_{\rm max}$}, without our completeness 
correction applied.  It agrees very well with Kepler \etal (2007---dots).  
The solid line is with our completeness correction applied, and represents 
the true local space density of white dwarfs.  The bottom panel shows the ratio 
of our two mass functions---\ie the cumulative completeness correction for each 
bin.  The small variation indicates that the completeness correction, while 
changing the overall normalization by roughly a factor of 2.2, has little effect 
on the shape of the MF.}
\label{wdmfcomp}
\end{figure}

Figure \ref{wdmfcomp} shows the WD mass function we derive for all stars with 
\Teff{ }$>12,000K$ and $g < 19.0$.  The red dashed line is the MF corrected
only by \Vmax---\ie before we apply our completeness correction.  It 
generally shows good agreement with the MF derived in \citet{Kepler07} 
(blue points), not surprising considering we use nearly the same 
data set and very similar WD models.  The small differences are
due to our use of slightly different sets of data and models, as well as
differing treatment of duplicate spectra, and can largely be considered 
statistical fluctuations.  We refer the interested reader to their paper 
for a more in depth analysis of the WDMF.

The solid black line in the upper panel shows our MF after correcting for 
the completeness of the spectroscopic sample.  This curve represents the true 
local space density of WDs per cubic parsec per ${M}_{\odot}$ interval.  
The bottom panel shows the total weight of each bin above the uncorrected 
MF---essentially the final completeness correction for each bin.  There is 
little small scale variation from bin to bin, and our completeness correction 
mainly has the effect of raising the normalization of the whole MF by a 
factor of $\sim2.2$.  In other words, the shape of the MF is not strongly 
affected by the completeness correction.

\begin{figure}[!tp]
\includegraphics[angle=270,width=\textwidth]{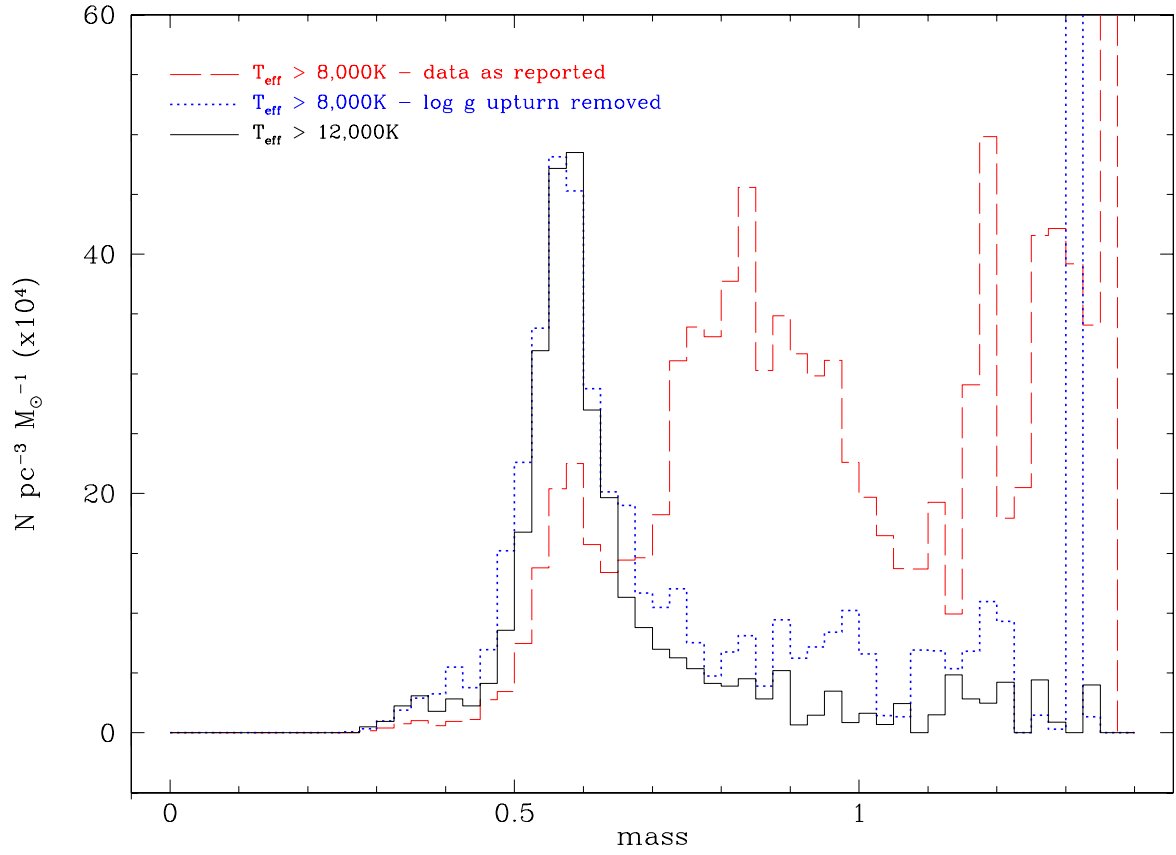}
\caption{White dwarf mass functions for WDs with ${T}_{\mathrm{eff}} >$ 
8,000K and $g < 19.0$ both with and without the upturn in log $g$ for cooler stars 
removed.  The solid line is the MF from Figure \ref{wdmfcomp} renormalized
for comparison purposes.}
\label{allwdmf}
\end{figure}

Figure \ref{allwdmf} is the WDMF for all stars down to 8,000K.  The dashed red
line is for the data as reported by \citet{Kepler07}, the dotted blue line is
after our correction for the upturn in log $g$.  The solid black line
is the WDMF for only those stars above 12,000K (\ie the same as Figure 
\ref{wdmfcomp}) renormalized to the same scale for comparison purposes.  
There are more high mass stars in general, and one spurriously large bin, 
but on the whole, our log $g$ correction recovers a reasonable mass distribution 
for stars cooler than 12,000K.

\begin{figure}[!tp]
\includegraphics[angle=270,width=\textwidth]{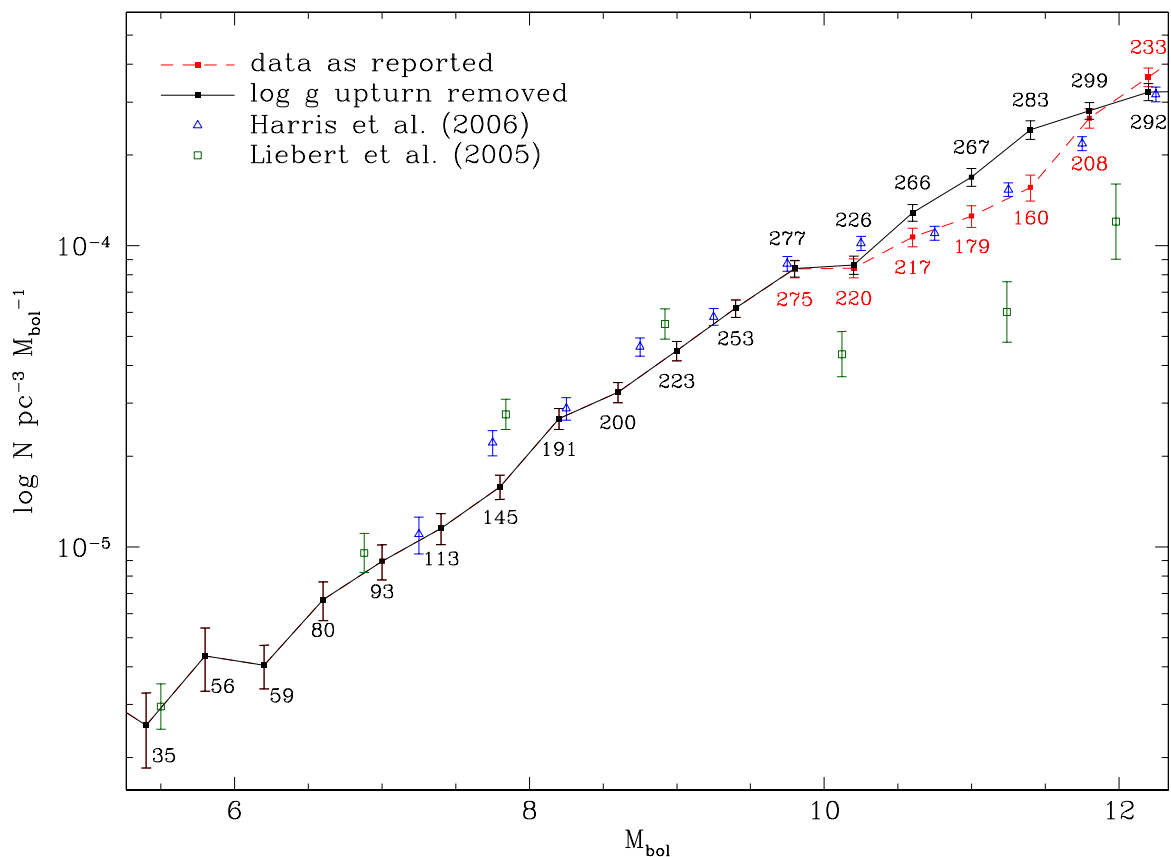}
\caption{LFs derived in this paper.  Removing the log $g$ upturn 
makes each affected star less massive, and therefore larger and brighter, 
pushing it to a more leftward ${M}_{bol}$ bin.  The results of \citet{Harris06} 
and \citet{Liebert05} are shown for comparison.}
\label{allwdlf}
\end{figure}

Figure \ref{allwdlf} shows the luminosity function we derive for all 
of the DA stars in our sample down to 7,000K for all stars with $g<19.5$.  
In red is the LF for the data as reported; in black is the LF for the 
data with the increase in log $g$ at low temperature removed.  The 
process of removing the excess log $g$ pushes stars to lower masses, 
making them larger and therefore brighter for the same \Teff.  In the 
range plotted, the black curve contains a total of 3,358 WDs, while 
the red contains 2,940.

\begin{figure}[!tp]
\includegraphics[angle=270,width=\textwidth]{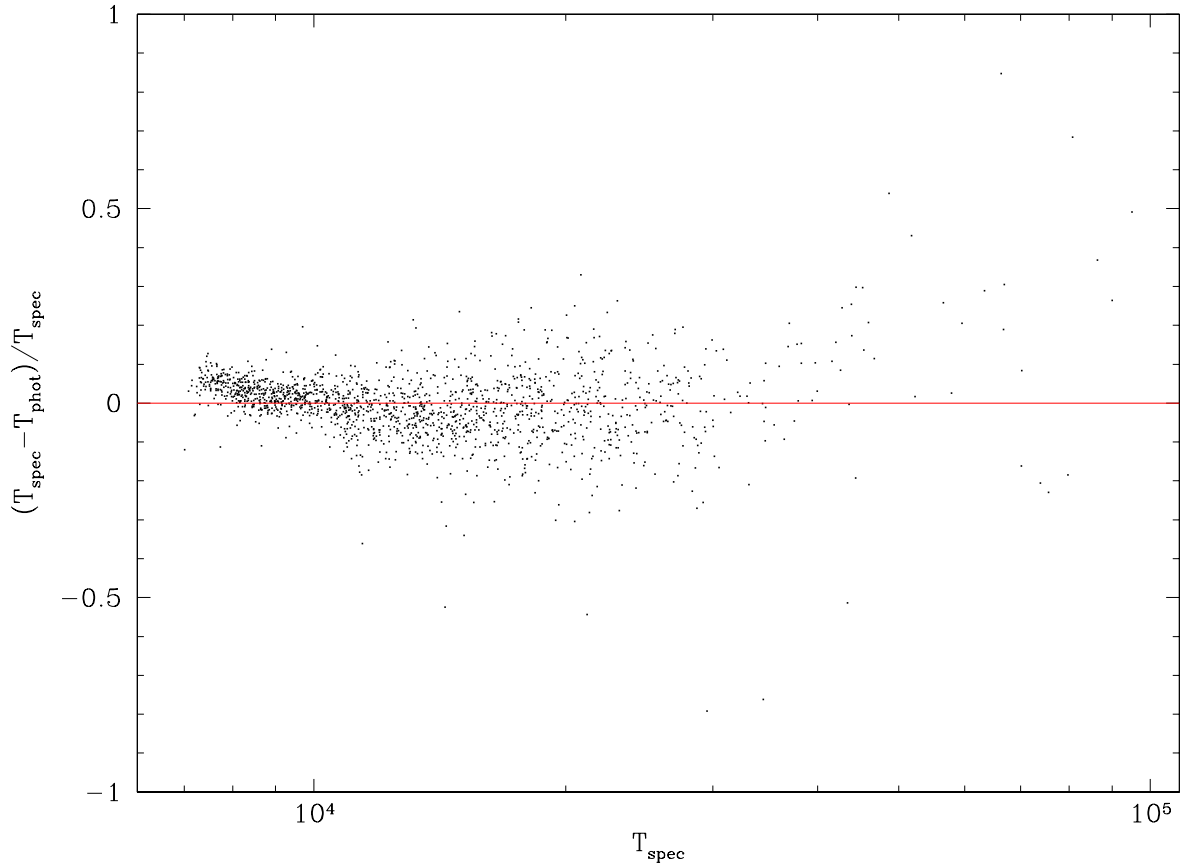}
\caption{A comparison of the spectrally and photometrically derived 
temperatures for the WDs common to the \citet{Harris06} and Eisenstein \etal 
(2006) samples.}
\label{Tcompfig}
\end{figure}

The lack of agreement between our best LF (black) and the \citet{Harris06}
luminosity function (blue) can be attributed, at least in part, to the 
differing assumptions used in creating the two LFs.  Harris \etal derived 
their temperatures by fitting Bergeron models to the photometry assuming 
a log $g$ of 8.0 for every star, a poor assumption for more than 30\% of WDs
\citep{Liebert05, Kepler07}.  The temperatures they derive are 
systematically different from the spectroscopic temperatures; Figure 
\ref{Tcompfig} shows the fractional difference between the spectroscopically 
and photometrically derived effective temperatures.  When we use the 
photometrically derived temperatures and set {log $g = 8.0$}, we recover 
the Harris \etal LF fairly well.

It should also be noted that the Harris \etal luminosity function is for 
WDs of $all$ types, whereas ours is comprised only of the DAs.  For 
each bin in the Harris \etal LF, we have used the full \citet{Eisenstein06} 
catalog to determine a rough DA fraction, and reduced the LF reported of 
Harris \etal accordingly.  This DA fraction---shown in Table 1---is in 
generally good agreement with previous works \citep{FLG}, but we have 
made no attempt to address selection biases in the Eisenstein \etal catalog.

One other source of the discrepancy between our results and Harris \etal
is due to our assumption that whatever causes the observed upturn in log $g$ 
in the cooler stars affects only the log $g$ determination and does not 
alter the spectroscopically derived \Teff.  As the effects of the
two parameters on the line profiles are interdependent, this assumption is
probably not valid.  The curves in Figure \ref{allwdlf} suggest that in 
addition to the excess \logg, the temperatures determined by line fitting
for the cooler stars are probably too high.  Ultimately, this area of the 
spectroscopic WDLF will remain uncertain until the problems with the model
atmospheres have been resolved.

The LF of \citet{Liebert05} shown in green in figure \ref{allwdlf} was 
compiled from a small dataset (348 DA white dwarfs) based on a survey done 
on photographic plates over 20 years ago on a 0.5m telescope.  In addition 
to low number statistics, the dataset suffers from a very difficult-to-quantify  
incompleteness on the faint end, which is probably responsible for the lack of 
agreement below ${M}_{bol}\sim9.5$.

\section{Conclusions}

Our eventual goal is to take advantage of the tremendous number of WDs 
spectroscopically observed by SDSS and studied by \citet{Eisenstein06} 
and others to create separate WD luminosity functions for two or more different 
ranges of mass.  This will effectively add a third dimension, currently 
unexplored, to observational WD luminosity functions.

In order to carry out this analysis, we must fully understand the manner 
in which white dwarfs were selected to receive spectra in SDSS.  By 
comparing the proper-motion selected sample of \citet{Harris06} with 
the spectroscopically determined WDs of \citet{Kleinman04} and 
\citet{Eisenstein06}, we have derived a WD selection probability 
over a range of parameters that includes nearly the entire useful range 
of $g-i$ color ($-1.0 < g-i < 0.2$) and apparent $g$ magnitude ($15 < g < 19.5$).

We have also presented additional arguments that the observed upturn in 
\logg{ }is an artifact of the model atmosphere line-fitting procedure, 
or---more likely---a problem with the line profiles themselves.  Since it may 
be some time before this problem is fully understood and addressed, we have 
implemented a procedure to remove the excess \logg{ }empirically and
shown that the mass function recovered for the stars cooler than 12,000K
reasonably agrees with the MF for the hotter stars, which in turn agrees
well with previous work.

Finally, we have presented the first WDLF for spectroscopically determined WDs
in the Fourth Data Release of the SDSS.  In addition to addressing the
issues of completeness and the observed log $g$ upturn in a more systematic
manner than previously attempted, our LF contains the largest sample of
spectroscopically determined WDs to date (3,358), more than six times the 
531 presented in \citet{Hu07}, and more than an order of magnitude
more than the 298 stars included in the LF of \citet{Liebert05}.

\acknowledgments
We would like to thank Scot Kleinman for providing unpublished fits of the 
DA white dwarfs; Hugh Harris and Mukremin Kilic for access to the data and 
code used to derive the \citet{Harris06} luminosity function, as well as 
much helpful advice; Barbara Canstanheira, Elizabeth Jeffery, Agnes Kim, 
Mike Montgomery, Fergal Mullally, and Kurtis Williams for many interesing 
and insightful discussions.

This material is based on work supported by the National Science
Foundation under grants AST 03-07315 and AST 06-07480.  This material
is partially based upon work supported by the National Aeronautics and
Space Administration under Grant No.\ NAG5-13070 issued through the
Office of Space Science.

\newpage

\clearpage

\begin{table}[!htbp]
\begin{center}
\begin{tabular}[c]{|c|c|}
\multicolumn{2}{p{8cm}}{\caption{The fraction of stars in 
\citet{Eisenstein06} listed as DA or DA\_auto.  Though they generally 
agree with previous results, they should be used with much caution, as they 
were calculated crudely and we have taken no care to correct for biases in 
the sample. We have employed them here simply to compare our DA-only luminosity 
function to previous work.}} \\
\hline
${M}_{bol}$ & DA Fraction\\
\hline
7.25 & 0.9338 \\
7.75 & 0.9243 \\
8.25 & 0.9246 \\
8.75 & 0.8980 \\
9.25 & 0.8433 \\
9.75 & 0.8146 \\
10.25 & 0.7958 \\
10.75 & 0.8158 \\
11.25 & 0.7957 \\
11.75 & 0.7721 \\
12.25 & 0.7985 \\
12.75 & 0.7976 \\
13.25 & 0.8173 \\
13.75 & 0.8009 \\
\hline
\end{tabular}
\end{center}
\label{DAfrac}
\end{table}


\begin{thebibliography}{50}
\expandafter\ifx\csname natexlab\endcsname\relax\def\natexlab#1{#1}\fi

\bibitem[{{Althaus} {et~al.}(2005){Althaus}, {Garc{\'{\i}}a-Berro}, {Isern}, \&
  {C{\'o}rsico}}]{Althaus05}
{Althaus}, L.~G., {Garc{\'{\i}}a-Berro}, E., {Isern}, J., \& {C{\'o}rsico},
  A.~H. 2005, \aap, 441, 689

\bibitem[{{Althaus} {et~al.}(2007){Althaus}, {Garc{\'{\i}}a-Berro}, {Isern},
  {C{\'o}rsico}, \& {Rohrmann}}]{Althaus07}
{Althaus}, L.~G., {Garc{\'{\i}}a-Berro}, E., {Isern}, J., {C{\'o}rsico}, A.~H.,
  \& {Rohrmann}, R.~D. 2007, \aap, 465, 249

\bibitem[{{Bergeron} {et~al.}(1995{\natexlab{a}}){Bergeron}, {Liebert}, \&
  {Fulbright}}]{Bergeron95b}
{Bergeron}, P., {Liebert}, J., \& {Fulbright}, M.~S. 1995{\natexlab{a}}, \apj,
  444, 810

\bibitem[{{Bergeron} {et~al.}(1995{\natexlab{b}}){Bergeron}, {Wesemael}, \&
  {Beauchamp}}]{Bergeron95a}
{Bergeron}, P., {Wesemael}, F., \& {Beauchamp}, A. 1995{\natexlab{b}}, \pasp,
  107, 1047

\bibitem[{{Bergeron} {et~al.}(1990){Bergeron}, {Wesemael}, {Fontaine}, \&
  {Liebert}}]{Bergeron90}
{Bergeron}, P., {Wesemael}, F., {Fontaine}, G., \& {Liebert}, J. 1990, \apjl,
  351, L21

\bibitem[{{Blanton} {et~al.}(2003)}]{Blanton03}
{Blanton}, M.~R. {et~al.} 2003, \aj, 125, 2276

\bibitem[{{Bradley}(1998)}]{Bradley98}
{Bradley}, P.~A. 1998, \apjs, 116, 307

\bibitem[{{Bradley}(2001)}]{Bradley01}
---. 2001, \apj, 552, 326

\bibitem[{{Bradley}(2006)}]{Bradley06}
---. 2006, Memorie della Societa Astronomica Italiana, 77, 437

\bibitem[{{Claver}(1995)}]{Claver95}
{Claver}, C.~F. 1995, PhD thesis, AA(THE UNIVERSITY OF TEXAS AT AUSTIN.)

\bibitem[{{Claver} {et~al.}(2001){Claver}, {Liebert}, {Bergeron}, \&
  {Koester}}]{Claver01}
{Claver}, C.~F., {Liebert}, J., {Bergeron}, P., \& {Koester}, D. 2001, \apj,
  563, 987

\bibitem[{{Eisenstein} {et~al.}(2006)}]{Eisenstein06}
{Eisenstein}, D.~J. {et~al.} 2006, \apjs, 167, 40

\bibitem[{{Engelbrecht} \& {Koester}(2007)}]{Engelbrecht06}
{Engelbrecht}, A. \& {Koester}, D. 2007, in press, Proceedings of the 15th
  European Workshop on White Dwarfs, Leicester 2006

\bibitem[{{Finley} {et~al.}(1997){Finley}, {Koester}, \& {Basri}}]{Finley97}
{Finley}, D.~S., {Koester}, D., \& {Basri}, G. 1997, \apj, 488, 375

\bibitem[{{Fleming} {et~al.}(1986){Fleming}, {Liebert}, \& {Green}}]{FLG}
{Fleming}, T.~A., {Liebert}, J., \& {Green}, R.~F. 1986, \apj, 308, 176

\bibitem[{{Fontaine} {et~al.}(2001){Fontaine}, {Brassard}, \&
  {Bergeron}}]{Fontaine01}
{Fontaine}, G., {Brassard}, P., \& {Bergeron}, P. 2001, \pasp, 113, 409

\bibitem[{{Geijo} {et~al.}(2006){Geijo}, {Torres}, {Isern}, \&
  {Garc{\'{\i}}a-Berro}}]{Geijo06}
{Geijo}, E.~M., {Torres}, S., {Isern}, J., \& {Garc{\'{\i}}a-Berro}, E. 2006,
  \mnras, 369, 1654

\bibitem[{{Green}(1980)}]{Green80}
{Green}, R.~F. 1980, \apj, 238, 685

\bibitem[{{Hansen} {et~al.}(2002)}]{Hansen02}
{Hansen}, B.~M.~S. {et~al.} 2002, \apjl, 574, L155

\bibitem[{{Hansen} {et~al.}(2004)}]{Hansen04}
---. 2004, \apjs, 155, 551

\bibitem[{{Hansen} {et~al.}(2007)}]{Hansen07}
---. 2007, ArXiv Astrophysics e-prints

\bibitem[{{Harris} {et~al.}(2003)}]{Harris03}
{Harris}, H.~C. {et~al.} 2003, \aj, 126, 1023

\bibitem[{{Harris} {et~al.}(2006)}]{Harris06}
---. 2006, \aj, 131, 571

\bibitem[{{Hu} {et~al.}(2007){Hu}, {Wu}, \& {Wu}}]{Hu07}
{Hu}, Q., {Wu}, C., \& {Wu}, X.-B. 2007, \aap, 466, 627

\bibitem[{{H{\"u}gelmeyer} {et~al.}(2006)}]{Hugelmeyer06}
{H{\"u}gelmeyer}, S.~D. {et~al.} 2006, \aap, 454, 617

\bibitem[{{Jeffery} {et~al.}(2007)}]{Jeffery07}
{Jeffery}, E.~J. {et~al.} 2007, \apj, 658, 391

\bibitem[{{Kepler} {et~al.}(2007)}]{Kepler07}
{Kepler}, S.~O. {et~al.} 2007, \mnras, 375, 1315

\bibitem[{{Kleinman} {et~al.}(2004)}]{Kleinman04}
{Kleinman}, S.~J. {et~al.} 2004, \apj, 607, 426

\bibitem[{{Knox} {et~al.}(1999){Knox}, {Hawkins}, \& {Hambly}}]{Knox99}
{Knox}, R.~A., {Hawkins}, M.~R.~S., \& {Hambly}, N.~C. 1999, \mnras, 306, 736

\bibitem[{{Koester} {et~al.}(2001)}]{Koester01}
{Koester}, D. {et~al.} 2001, \aap, 378, 556

\bibitem[{{Krzesi{\'n}ski} {et~al.}(2004)}]{Krzesinski04}
{Krzesi{\'n}ski}, J. {et~al.} 2004, \aap, 417, 1093

\bibitem[{{Leggett} {et~al.}(1998){Leggett}, {Ruiz}, \& {Bergeron}}]{Leggett98}
{Leggett}, S.~K., {Ruiz}, M.~T., \& {Bergeron}, P. 1998, \apj, 497, 294

\bibitem[{{Liebert} {et~al.}(2005){Liebert}, {Bergeron}, \&
  {Holberg}}]{Liebert05}
{Liebert}, J., {Bergeron}, P., \& {Holberg}, J.~B. 2005, \apjs, 156, 47

\bibitem[{{Liebert} {et~al.}(1979){Liebert}, {Dahn}, {Gresham}, \&
  {Strittmatter}}]{Liebert79}
{Liebert}, J., {Dahn}, C.~C., {Gresham}, M., \& {Strittmatter}, P.~A. 1979,
  \apj, 233, 226

\bibitem[{{Liebert} {et~al.}(1988){Liebert}, {Dahn}, \& {Monet}}]{LDM}
{Liebert}, J., {Dahn}, C.~C., \& {Monet}, D.~G. 1988, \apj, 332, 891

\bibitem[{{Luyten}(1958)}]{Luyten58}
{Luyten}, W.~J. 1958, \emph{On the Frequency of White Dwarfs in Space}
  (Minneapolis: University of Minnesota Observatory)

\bibitem[{{Munn} {et~al.}(2004)}]{Munn04}
{Munn}, J.~A. {et~al.} 2004, \aj, 127, 3034

\bibitem[{{Richards} {et~al.}(2002)}]{Richards02}
{Richards}, G.~T. {et~al.} 2002, \aj, 123, 2945

\bibitem[{{Richer} {et~al.}(1998){Richer}, {Fahlman}, {Rosvick}, \&
  {Ibata}}]{Richer98}
{Richer}, H.~B., {Fahlman}, G.~G., {Rosvick}, J., \& {Ibata}, R. 1998, \apjl,
  504, L91+

\bibitem[{{Schlegel} {et~al.}(1998){Schlegel}, {Finkbeiner}, \&
  {Davis}}]{Schlegel98}
{Schlegel}, D.~J., {Finkbeiner}, D.~P., \& {Davis}, M. 1998, \apj, 500, 525

\bibitem[{{Schmidt}(1968)}]{Schmidt68}
{Schmidt}, M. 1968, \apj, 151, 393

\bibitem[{{Stoughton} {et~al.}(2002)}]{Stoughton02}
{Stoughton}, C. {et~al.} 2002, \aj, 123, 485

\bibitem[{{von Hippel} {et~al.}(1995){von Hippel}, {Gilmore}, \&
  {Jones}}]{vonHippel95}
{von Hippel}, T., {Gilmore}, G., \& {Jones}, D.~H.~P. 1995, \mnras, 273, L39

\bibitem[{{von Hippel} {et~al.}(2006)}]{vonHippel06}
{von Hippel}, T. {et~al.} 2006, \apj, 645, 1436

\bibitem[{{Weidemann}(1967)}]{Weidemann67}
{Weidemann}, V. 1967, Zeitschrift fur Astrophysik, 67, 286

\bibitem[{{Winget} {et~al.}(1987)}]{Winget87}
{Winget}, D.~E. {et~al.} 1987, \apjl, 315, L77

\bibitem[{{Wood}(1992)}]{Wood92}
{Wood}, M.~A. 1992, \apj, 386, 539

\bibitem[{{Wood}(1995)}]{Wood95}
{Wood}, M.~A. 1995, in Lecture Notes in Physics, Berlin Springer Verlag, Vol.
  443, White Dwarfs, ed. D.~{Koester} \& K.~{Werner}, 41

\bibitem[{{Wood} \& {Oswalt}(1998)}]{Wood98}
{Wood}, M.~A. \& {Oswalt}, T.~D. 1998, \apj, 497, 870

\bibitem[{{York} {et~al.}(2000)}]{York00}
{York}, D.~G. {et~al.} 2000, \aj, 120, 1579

\end{thebibliography}
\end{document}